\documentclass[a4paper]{article}
\usepackage{amsfonts}
\usepackage{amsmath,amssymb,amsthm}
\usepackage{latexsym}
\usepackage[english]{babel}
\usepackage{graphicx}
\begin{document}
\title{Q-balls  in  the regularized signum-Gordon model}
\author{Jakub Lis \thanks{lis@th.if.uj.edu.pl}\\
Institute of Physics, Jagiellonian University\\
\small{Reymonta 4, 30-059 Cracow, Poland}
}
\maketitle

\begin{abstract}
The regularized signum-Gordon potential has a  smooth minimum and is linear in the modulus of the field value for higher amplitudes. The Q-ball solutions in this model are investigated.  Their existence for charges large enough is demonstrated. In three dimensions  numerical solutions are presented and the absolute stability of large Q-balls is proved. It is also shown, that the solutions of the regularized model approach uniformly  the solution of the un\-re\-gu\-la\-rized signum-Gordon model. From the stability of Q-balls in the regularized model follows the stability of the solutions in the original theory.
\end{abstract}

\section{Introduction}
In the seminal paper \cite{coleman} S. Coleman addressed the following problem. Consider a field theory with a symmetry  in the internal field space. Then, due to the Noether theorem there is a charge $Q$ in the system, a quantity constant in time. A legitimate problem is then what solution minimizes the energy~$E$ for a given~$Q$. S. Coleman  managed to answer this question for a class of ``acceptable" field potentials and gave a recipe how to find the relevant solutions. He dubbed them Q-balls. The time dependence of Q-balls is confined to the space of the field symmetry, so that the energy and charge density do not evolve in time (actually up to Lorentz boosts). The space distribution of the field is given by a spherically symmetric, positive and monotone (as a function of the radial coordinate) decreasing function. These solutions are of physical importance and much attention has been paid to them, see~\cite{balls}. The Q-ball Ansatz may be also useful in  models spoiling the prerequisites given by Coleman (see e.g. \cite{ar}, \cite{vir}). Then, the status of the solutions is not clear; they may be just unstable configurations or absolutely stable solutions. In this paper we show, that for the scalar complex field with the ``unacceptable" potential $V(\psi)\sim\sqrt{|\psi|^{2}+\epsilon}-\epsilon$ such solutions may fall into the second  category. In what follows we refer to this potential as regularized signum-Gordon one. This name traces back to  original motivation. 

Recently considerable work has been done in exploring the signum-Gordon model (\cite{ar},\cite{ar2}), where the field potential $V(\psi)\sim|\psi|$. One of the intriguing characteristics of such field theory is   the absence of the linear regime ("infinite" mass). It results in the compactness of solutions, the Q-balls described in \cite{ar} are paradigmatic. In this reference it is also pointed out, that the signum-Gordon model may be regarded as a limiting case for the regularized one when $\epsilon \rightarrow 0$. The  considerations presented below support the suggestion: the Q-balls in regularized theory tend uniformly with $\epsilon$ to these ones known from the "sharp" potential. Consequently,  the global quantities: charge and energy approach the relation found in ref. \cite{ar}.

The paper is organized as follows. The next section is devoted to Q-balls in the regularized signum-Gordon model. Numerical results in the case of the three spatial dimensions are reviewed in section \ref{numerics}. In the section \ref{signum} we examine briefly the equation motivated by the Q-ball Ansatz in the signum-Gordon model in any number of space dimensions. To this end, we recall and supplement the results presented in ref. \cite{ar}.
In the section \ref{limes}  the limit  $\epsilon\rightarrow 0$ is taken for solutions of the regularized model. Finally, in  section \ref{proof}  we adapt the S. Coleman's proof of Q-ball stability for the regularized model (also in three dimensions). An immediate consequence of the stability of the Q-balls in the regularized model is the stability of the solutions in the original model. In the last section  we summarize the results and discuss  some open problems. 

\section{The regularized model}\label{sect:regul}

The Lagrangian defining the theory of interest has the form
\begin{equation}\label{lag}
L=\partial_{\mu}\Phi \partial^{\mu}\bar{\Phi}-\lambda \sqrt{\epsilon^{2}+\Phi\bar{\Phi}}+\lambda \epsilon,
\end{equation}
where $\Phi$ and $\bar{\Phi}$ denote the scalar field and its complex
conjugation, $\epsilon$ and $\lambda$ are positive real numbers.
The Lagrangian respects  Lorentz symmetry. The space dimension
$n$ does not need to be specified now. A global change of the
field phase  does not affect the Lagrangian giving rise to the
Noether charge 
\begin{equation}\label{char}
Q=\frac{1}{2i} \int d^{n}x\left[\partial_{t} \Phi\bar{\Phi}- \Phi\partial_{t}\bar{\Phi}\right].
\end{equation}
This makes the Coleman's question about field configuration minimizing the energy $E$ for a given charge $Q$ relevant. Although the theory (\ref{lag}) is not an ``acceptable"  one (a  discussion of the acceptability is postponed to section \ref{proof}), we  plug the Q-ball Ansatz  into the field equations. The Q-ball Ansatz for the complex scalar field has the form
\begin{equation}\label{Ansatz}
\Phi(t,\vec{x}) = F(r) \exp\left(i\omega t\right),
 \end{equation}
where  $F$ is a real valued function of the radial coordinate $r$ and $\omega>0$.
After rescaling of the radial variable $y= \omega r$ and the profile function $f_{\delta}(y)=\frac{2 \omega^{2}}{\lambda}F$ the following equation is obtained
\begin{equation}\label{regul}
f_{\delta}''+\frac{n-1}{y}f_{\delta}'+f_{\delta}=\frac{f_{\delta}}{\sqrt{\delta^{2}+f_{\delta}^{2}}}.
\end{equation}
The $'$ stands for derivative with respect to $y$ and $\delta=2
\omega^{2}\epsilon/\lambda $. The physically meaningful solution obeys the
conditions: $f'_{\delta}(0)=0$ and $f_{\delta}(\infty)=0$. Such a solution of the above equation is denoted as $\hat{f}_{\delta}$.

Now, we argue that the equation (\ref{regul}) has at least one meaningful solution. 
The above equation may be
interpreted in terms of classical mechanics - it corresponds to a
point  particle moving in a potential
$f_{\delta}^{2}/2-\sqrt{f_{\delta}^{2}+\delta^{2}}$ and subject to
the time  dependent  friction - $y$ is regarded here as time. The potential changes qualitatively
for $\delta=1$. If $\delta>1$, it has one
global minimum and  only the trivial solution ($f_{\delta}=0$) satisfies the
equation (\ref{regul}) and boundary conditions.  For
$0<\delta < 1$ the  potential has a local maximum at $f=0$
and two  symmetric global  minima for $f=\pm \sqrt{1-\delta^{2}}$.  In
this range of the parameter $\delta$ the equation (\ref{regul}) admits
nontrivial solutions. The heuristic reasoning for their existence
uses the continuity argument as follows: it is possible to find such $f_{\delta}(0)$ that the particle cannot pass the local
maximum and oscillates forever around one of the minima. However,
for another $f_{\delta}(0)$, larger than the previous one, the particle
may cross the local maximum and dip on its other side. In between
the two families of solutions the sought after solution is expected.
The fact of the existence of the two families is  demonstrated in
the section \ref{r2s}. The exact solutions of the  equation
(\ref{regul}) are not known, so we have to resort to the numerics.
The results are presented in the  section \ref{numerics}. \\
The mechanical analogy is useful to demonstrate that the one-dimensional model has the relevant solutions. In this case there is no friction. Hence, the equation gains an integral of motion corresponding to the mechanical energy
\begin{displaymath}
E_{mech}=\frac{1}{2}{f'_{\delta}}^{2}+\frac{1}{2}f_{\delta}^{2}+\delta-\sqrt{\delta^{2}+f_{\delta}^{2}}.
\end{displaymath}
The Q-ball solution emerges  for $E_{mech}=0$, the inverse function has then the form
\begin{displaymath}
y(f_{\delta})=\int_{2\sqrt{1-\delta}}^{f_{\delta}}\frac{dx}{\sqrt{2\sqrt{x^{2}+\delta^{2}}-x^{2}-2\delta}}.
\end{displaymath}
A detailed analysis of the above formula  reveals the relation between the \mbox{Q-balls} in the regularized and the ``sharp" model. The same may be  achieved  with methods presented in section \ref{limes} (some obvious modifications are in order then). In the sequel we will no more return to the one-dimensional case.

Except for $n=2$, we can give a more reliable argument for existence of the required solution. The argument follows from a theorem proved in ref.~\cite{coleman2}. That theorem states that the equation 
\begin{equation}\label{pomoc1}
\Delta \psi=\frac{dU(\psi)}{d\psi},
\end{equation}
has at least one spherically symmetric positive, monotone and vanishing in infinity solution. What is more, the integrability of the terms $U(\psi)$ and $(\nabla\psi)^{2}$ is also granted. In the above equation  $\psi$ denotes a real valued function and $\Delta$ stands for a Laplacian in $n>2$ dimensions. It holds if $U$ satisfies four conditions:
\begin{enumerate}
\item $U$ is continuously differentiable for all $\psi$;
\item $U(0)=U'(0)=0$;
\item $U$ is somewhere negative;
\item There exist positive numbers $a$, $b$, $\alpha$ and $\beta$ such that $\alpha<\beta<2n/(n-2)$ and 
\begin{displaymath}
U\geq a|\psi|^{\alpha}-b|\psi|^{\beta}.
\end{displaymath}
\end{enumerate}
Let us  consider
\begin{displaymath}
U(\psi)=\left(\sqrt{\psi^{2}+\delta^{2}}-\delta\right)-\frac{1}{2}\psi^{2}.
\end{displaymath}
This $U$ satisfies the above requirements (e.g. for $\delta<0.95$ one can take \mbox{$\alpha=2$}, \mbox{$\beta=3$}, \mbox{$a=(1-\delta)/2$}, \mbox{$b=4(1-\sqrt{\delta})\delta^{-1}$}).  Thus, the theorem applies. It is clear, that the solution described in the theorem corresponds to the Q-ball solution of the eq. (\ref{regul}). Alas, in two spatial  dimensions we have nothing but the heuristic argument.

\section{Numerical results} \label{numerics}
\begin{figure}
\centering
\input{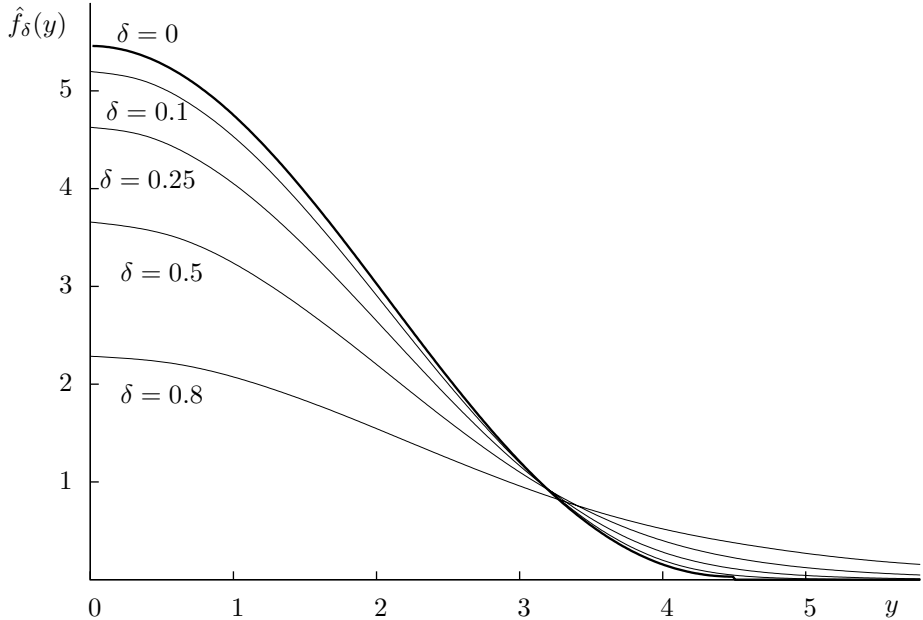}
\caption{Profile function for various $\delta$ values. The solution of signum Gordon model is marked as $\delta=0$.}
\label{kszt}
\end{figure}

The  numerical analysis of the regularized signum-Gordon model is
done for $n=3$ space dimensions. Some  profile functions
$f_{\delta}$ for various $\delta$ are depicted in  fig.
~\ref{kszt}. The relevant solution of the original
signum-Gordon model  is also plotted in this figure. Such a
presentation  supports the supposition that the solution of the
``sharp" potential is a limiting case for the solutions of  the regularized problem.

\begin{figure}
\centering
\includegraphics[width=0.6\textwidth,height=0.3\textheight]{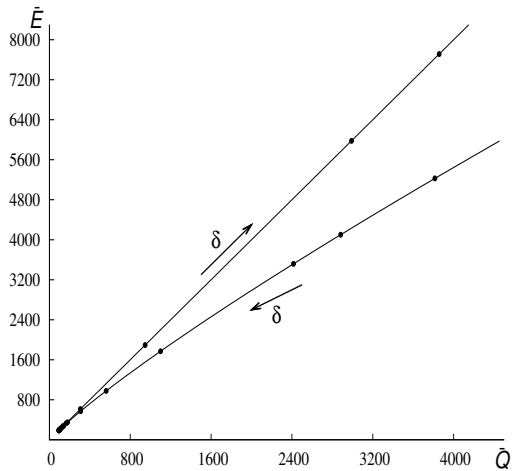}
\caption{Relation $\bar{E}(\bar{Q})$ for the regularized signum-Gordon model. For  a given charge there exist two different Q-balls with different energies resulting in a cusp on the graph. The arrows show how $\delta$ changes along the curve.}
\label{dqen}
\end{figure}

 The relations between the charge, energy and the parameter $\delta$ are most inte\-resting from the physical viewpoint.
Plugging the Ansatz~(\ref{Ansatz}) into the definition of the charge (\ref{char}) we obtain
\begin{equation}\label{ladun}
Q=\frac{\pi}{\lambda}\left(\frac{ 2\epsilon}{\delta}\right)^{3}\int \hat{f}_{\delta}^{2}r^{2} \ dr=\frac{\left( 2\epsilon\right)^{3}}{\lambda}\bar{Q}(\delta).
\end{equation}
The energy is given by the formula
\begin{equation}\label{ener}
E=\frac{\pi}{\sqrt{\lambda}} \left(\frac{2 \epsilon}{\delta}\right)^{5/2}\int dr \ r^{2} \left[ (\hat{f}_{\delta}{'})^{2} + \hat{f}_{\delta}^{2} + 2\left(\sqrt{\hat{f}_{\delta}^{2}+\delta^{2}}-\delta\right)\right]=\frac{\left(2 \epsilon\right)^{5/2}}{\sqrt{\lambda}} \bar{E}(\delta).
\end{equation}
$\bar{Q}$ and $\bar{E}$ are  functions of the
parameter $\delta$ only.  The relation $\bar{E}(\bar{Q})$  is shown in fig. \ref{dqen}.

\begin{figure}
\centering
\includegraphics[width=0.6\textwidth,height=0.3\textheight]{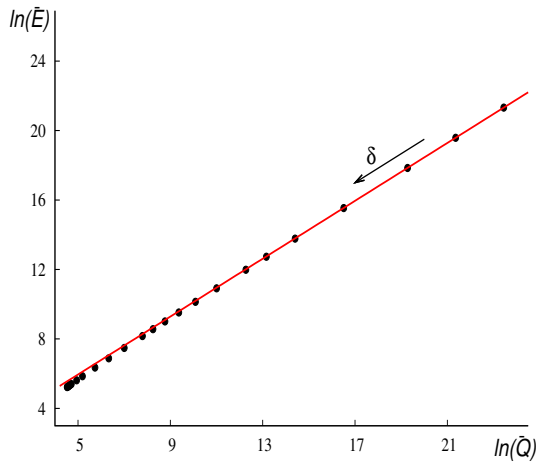}
\caption{The points come from the numerics of regularized model. The solid line illustrates the relation (\ref{sigEQ}). The agreement is excellent.}
\label{deqas}
\end{figure}

Quite a general feature in  theories with Q-balls is the existence
of a solution with minimal possible  charge and  energy value.
This is not the case of the ``sharp" signum-Gordon model, where
\begin{equation}\label{sigEQ}
E=\left(\frac{5\pi}{6}\right)^{1/6}\frac{6 \lambda^{1/3}}{5}Q^{5/6}
\end{equation}
for any charge $Q>0$.  The $E(Q)$ relation in the regularized model inherits  both from the ordinary models and the signum-Gordon one. As in most models there is a Q-ball with the smallest possible charge and energy (the corresponding solution is found for $\delta=0.96$, see tab. \ref{tab}). Two branches of the relation $E(Q)$ originate from the point corresponding to this solution, see fig. \ref{dqen}. The branch corresponding to  larger $\delta$'s has larger energy values.  Physically more favorable are solutions with smaller $\delta$ - they  may be absolutely stable. This lower branch of solutions  reproduces the power dependence $E(Q)$ known from the signum-Gordon model, see fig. \ref{deqas}. The energy and charge are smooth functions of  $0<\delta<1$.

The corner stone for the Q-ball theory is their absolute stability. In the section \ref{proof} it is shown, that the criterion ensuring this reads $E<Q \sqrt{2\lambda/\epsilon}$, where the proportionality coefficient between $E$ and $Q$ is the mass parameter of the theory. In case of our model the inequality may be written in a dimensionless form
\begin{equation}
2>\frac{\bar{E}}{\bar{Q}}.
\end{equation}
The numerics indicates, that  the inequality is
violated if $\delta>0.91$. It means, that almost all solutions from the lower branch are absolutely stable. The solutions lying on the upper branch cannot be absolutely stable, however they seem to be very close to the relation $\bar{E}=\sqrt{2\lambda/\epsilon}\bar{Q}$.  There are two more types of stability of Q-balls (see \cite{tsumagari}): linear (classical) stability and stability against fission. The classical stability is granted, if 
\begin{displaymath}
\frac{\omega}{Q}\frac{dQ}{d\omega}\leq 0,
\end{displaymath}
where $\omega$ denotes  the same quantity as in  (\ref{Ansatz}). In case of our model, where $\omega\sim\sqrt{\delta}$, the solutions from the lower branch of the relation $E(Q)$ satisfy the condition. It turns out, rather unexpectedly, that the condition for the stability against fission coincides with this for linear stability. Thus, the solutions from the lower branch are physically relevant, although not all of them are absolutely stable. Some data useful for numerical analysis   are given in  table \ref{tab}.

\begin{table}
\centering
\begin{tabular}{|c|c|c|c|}
\hline
$\delta$ & $f_{\delta}(0)$ & \small{$\bar{Q}$} & \small{$\bar{E}$}\\
\hline
0.9999&0.0613&947.16&1894.41\\
\hline
0.999&0.1928&306.32&  612.95\\
\hline
0.99&0.5844&117.44& 235.75\\
\hline
0.97&0.9629&93.62&  188.49\\
\hline
0.96&1.0948&92.62   &186.53\\
\hline
0.95&1.2094&93.59&  188.43\\
\hline
0.8&    2.2862&179.77 & 348.21\\
\hline
0.5&    3.6589&1100.09 & 1770.26\\
\hline
0.25&4.6276&11668.8&    13653.2\\
\hline
0.1&5.1988&213385&  160261\\
\hline
\end{tabular}
\caption{Data of some exemplary solutions in the regularized model}
\label{tab}
\end{table}

\section{The signum-Gordon model}\label{signum}

The defining feature of the signum-Gordon model is the field potential term in the Lagrangian given by $\lambda |\Phi|$. We  plug the Q-ball Ansatz (\ref{Ansatz}) into the equation of motion, rescale the radial coordinate $y= \omega r$ and introduce  $f(y)=\frac{2 \omega^{2}}{\lambda}F$ as previously ($f$ without $\delta$ in subscript relates to the signum-Gordon model). This leads to the equation \cite{ar}
\begin{equation}\label{ostre}
f''+\frac{n-1}{y}f'+f=sign(f).
\end{equation}
Due to the symmetry  $f\rightarrow -f$ we can
consider only  the solutions with $f(0)>0$. Then, the above
equation is a linear equation with the source term equal to unity.
The homogeneous part is solved with the substitution
$f(y)=y^{-\alpha}R(y)$, where $\alpha=(n-2)/2$. In this way  the
equation transforms into the Bessel's differential equation of
order $\alpha$. Two linearly independent solutions of  (\ref{ostre})
$u_{1}$ and $u_{2}$ may be written in terms of the Bessel
functions of the first $J_{\alpha}(y)$ and the second
$Y_{\alpha}(y)$ kind \cite{podr}:
\begin{equation}
\begin{array}{lr}
u_{1}(y)=y^{-\alpha}J_{\alpha}(y),& u_{2}(y)=y^{-\alpha}Y_{\alpha}(y).\\
\end{array}
\end{equation}
For small values of $y$ the first solution behaves like
$u_{1}\approx a-by^{2}$  with $a,b>0$. On a larger scale it is
oscillating function with decreasing amplitude. $u_{2}$ behaves in
the vicinity of the origin like  $y^{-2\alpha}$. Thus, the
solution of the equation (\ref{ostre}) obeying the conditions
$f(0)>0$ and $f'(0)=0$ has the following form:
\begin{equation} \label{rozw}
f(y)=\frac{f(0)-1}{u_{1}(0)}u_{1}(y)+1.
\end{equation}
Strictly speaking, this function solves  (\ref{ostre})  as long as
$f(y)>0$. It has a simple structure:  the function $u_1$ is
rescaled and than shifted by the term $+1$. Thus, the positions of
the extrema of $f$ do not depend on the starting point $f(0)$. If
$f(0)>1$ (as is supposed in the sequel) the first extremum is
minimum. We denote the argument of this minimum with $y_0$ and note
that $u_{1}(y_{0})\neq0$. Let us also define
\begin{equation}
f_{0}=1-\frac{u_{1}(0)}{u_{1}(y_{0})},
\end{equation}
and point out that if $f(0)=f_{0}$, then $f(y_{0})=0$.  If
$f(0)<f_{0}$, $f$ is valid solution for all arguments as it stays
positive for all $y$. For $f(0)>f_0$ the value of $f$ at the first minimum is
negative. Thus, for some $y<y_0$  the function $f$ changes its
sign and ceases to solve  (\ref{ostre}). Following the solution
with $f(0)=f_{0}$ an ambiguity is encountered for $y=y_{0}$. The
equation admits three ways of continuation for $y>y_{0}$: a valid
solution may follow the r.h.s. of (\ref{rozw}) with plus or minus
sign or may be set to zero. It is our choice motivated by the
field theoretical context to stick to the last option. The
resulting  function composed of two pieces corresponds to the
Q-ball profile function and is further denoted as $\hat{f}$. We
will show that the regularization of the potential supports the
choice.

\section{The limit $\delta\rightarrow 0$}
\label{limes}
The numerical results from the  section~\ref{numerics} suggest, that the
Q-balls in the regularized model approach the solution of the
signum-Gordon model. Now, we can give some analytical arguments for this. The spatial dimension $n>1$ does not need  to be
specified in what follows. First, we will show that for profile functions in the regularized model $\hat{f}_{\delta}(0) \rightarrow f_{0}$ when $\delta$ tends to zero. Then, we will find an upper bound for the modulus of the difference between the solution of the model with $\delta>0$ and the one with $\delta=0$. As the bound tends to zero, the solutions of the regularized models approach  uniformly the solution characterized in sect.~\ref{signum}.

\subsection{General setting}\label{2fam}
Now we are in a position to explore the solutions of the equation (\ref{regul}) for small values of the parameter $\delta$. To deal with the limit we set the notation and give some general estimates used later. \\
Let us consider the difference between solutions of  (\ref{regul}) and (\ref{ostre})
\begin{equation}
\eta(y)=f(y)-f_{\delta}(y).
\end{equation}
The pertinent solutions obey the same initial conditions $f_{\delta}(0)=f(0)$ and $f_{\delta}'(0)=f'(0)=0$. Let us make it clear, that at this stage we investigate solutions of the equations that can spoil the condition $f(\infty)=f_{\delta}(\infty)=0$. As long as $sign(f(y))=+1$ the  differential equation holds
\begin{equation}\label{etaeq}
\eta''+\frac{n-1}{y}\eta'+\eta=\frac{\delta^{2}}{\left(\sqrt{\delta^{2}+f_{\delta}^{2}}+f_{\delta}\right) \sqrt{\delta^{2}+f_{\delta}^{2}}}.
\end{equation}
It is supplemented with the initial conditions $\eta(0)=0$ and
$\eta'(0)=0$. For notational convenience let us call the r.h.s. in
the above equation $\varphi(f_{\delta}(y))$ or  shorter
$\varphi(y)$. $\varphi$  seen as a function of positive
$f_{\delta}$ is a positive and monotone decreasing function. Some algebra makes evident that
\mbox{$\varphi(f_{\delta})<\delta^{2/3}/g$} (with $g$ being a
positive constant) as long as
\begin{equation}\label{estimd}
f_{\delta}>\delta^{2/3}\sqrt{\frac{g}{2}}\frac{1-\frac{\delta^{2/3}}{g}} {\sqrt{1-\frac{\delta^{2/3}}{2g}}} =\delta^{2/3}\sqrt{\frac{g}{2}}+o(\delta).
\end{equation}
The equation (\ref{etaeq}) is an inhomogeneous linear equation. The homogeneous part is the same as in (\ref{ostre}). Then, using $u_1$ and $u_2$ the  solution may be written in the form
\begin{equation}\label{rozw1}
\eta(y)= \int_{0}^{y}G(y,s)s^{n-1}\varphi(s)ds,
\end{equation}
where $G(y,s)$ is a Green function (see \cite{podr}) and does not
depend on the pa\-ra\-meter~$\delta$. It has the form
\begin{displaymath}
G(y,s)=\frac{u_{1}(s)u_{2}(y)-u_{2}(y)u_{1}(s)}{y^{n-1}\left(u'_{2}(y)u_{1}(y)-u'_{1}(y)u_{2}(y)\right)}.
\end{displaymath}
\emph{A priori} any combination of the functions $u_{1}$ and $u_{2}$ could be added to the solution (\ref{rozw1}), but the  boundary conditions exclude such terms. The integral does not give rise to any ambiguity or difficulty for $y\rightarrow 0$. The above form of $\eta$ gives the following bound
\begin{equation}\label{estim0}
|\eta(y)|\leq \max_{s\in(0,y)}\{\varphi(s)\}\int_{0}^{y}|G(y,s)s^{n-1}|ds.
\end{equation}
Hence, for all $y<y_{0}+1$ the  inequality holds
\begin{equation}\label{estim}
|\eta(y)|\leq \max_{s\in(0,y)}\{\varphi(s)\}\int_{0}^{y_{0}+1}|G(y,s)s^{n-1}|ds=g_{1} \max_{s\in(0,y)}\{\varphi(s)\},
\end{equation}
where the last equality defines $g_1$. In order to get another helpful
observation it is convenient to rewrite the equation (\ref{etaeq})
in the form
\begin{displaymath}
y^{n-1}\eta'(y)=\int_{0}^{y} s^{n-1}\left(\varphi(s)-\eta(s)\right)ds.
\end{displaymath}
Together with (\ref{estim})  this gives the bound on the derivative $\eta'(y)$ for all $y<y_{0}+1$:
\begin{equation}\label{estimder}
|\eta'(y)|\leq y^{1-n}\int_{0}^{y} s^{n-1}\left(|\varphi(s)|+|\eta(s)|\right)ds\leq g_{2}\max_{s\in(0,y)}\{\varphi(s)\},
\end{equation}
where $g_{2}=(y_{0}+1)(1+g_{1})/n$.\\

\subsection{The limit $f_{\delta}(0)$ with $\delta\rightarrow 0$} \label{r2s}
Now, we  demonstrate that  a solution with $f_{\delta}(0)$ far from $f_{0}$ cannot be the Q-ball profile function - it either has a minimum or changes its sign. First we deal with the solutions $f_{\delta}(0)=f(0)=f_{0}-\xi$, $\xi>0$. 
It is shown in the section~\ref{signum}, that then the equation (\ref{etaeq}) and the solution (\ref{rozw1}) hold for any argument $y$. Assume $\eta(y)$ does not tend to zero for $0<y<y_{0}+1$ when   $\delta$ gets smaller and smaller. The inequality (\ref{estim}) makes clear that it may be true only if  $\varphi(f_{\delta})\geq\delta^{2/3}/g_1$ on a finite segment. Equivalently, it means that $f_{\delta}<\delta^{2/3}\sqrt{g_{1}/2}$ on this segment as follows from (\ref{estimd}).   For continuity reason $f_{\delta}$  has to  reach  this value for the first time at a point $y_{s}$. At this point the relation holds 
\begin{equation}
|f(y_{s})|\leq|f_{\delta}(y_{s})|+ |\eta(y_{s})|\leq \delta^{2/3}\sqrt{\frac{g_{1}}{2}}+\delta^{2/3}.
\end{equation}
This inequality may be combined with an other one: $f(y_{0})\leq f(y_{s})$, what  restricts the initial conditions allowing the reasoning
\begin{equation}
\label{estimx}
0\leq \xi\leq \left|\frac{u_{1}(0)}{u_{1}(y_{0})}\right|\delta^{2/3}\left(\sqrt{\frac{g_{1}}{2}}+1\right).
\end{equation}
If this inequality is spoiled, $|\eta|$ is smaller then
$\delta^{2/3}/ g_{1}$ on the whole segment. From this we can infer the 
existence of a  minimum of $f_{\delta}$; it suffices that the function $f$ takes
a value bigger than $f(y_{0})+\delta^{2/3}$ twice in the segment (eventually one can consider a larger segment instead of the
arbitrarily taken $y_{0}+1$). Thus, for $\delta$ small enough, the solution interpreted as
Q-ball cannot start with $f_{\delta}(0)<f_{0}-\delta^{2/3}(\sqrt{g_{1}/2}+1)$.

Let us analyze  the case $f_{\delta}(0)=f(0)=f_{0}+\xi$. It is argued in the
section~\ref{signum}, that there exists  a point
$y_{1}<y_{0}$ for which $f(y_{1})=0$ and $f'(y_{1})<0$. Consider another point for which
$f_{\delta}(y_{z})=\delta^{2/3}g_{1}^{1/3}$. It means, that
$|\eta(y_{z})|<\delta^{2/3}g_{1}^{1/3}/2$. Such choice of the
function value ensures that  $0<f(y_{z})$. To see, that the
function $f_{\delta}$ reaches the requested value assume the
contrary: $f_\delta$ does it not. As it is a continuous function, it
is always bigger than this (i.e. $\delta^{2/3}g_{1}^{1/3}$) value.  Then, $\eta$ is small enough to
ensure that $f\neq0$ for any value $0<y<y_{0}$, what is false. Now,
we can show that $f_{\delta}$ changes its sign if
$\delta$ is sufficiently small.  To this end we make use of
(\ref{estimder}) to get
\begin{equation}\label{esA}
f'(y_{z})-\frac{g_{2}g_{1}^{-2/3}}{2}\delta ^{2/3}<f'_{\delta}(y_{z})<f'(y_{z})+\frac{g_{2}g_{1}^{-2/3}}{2}\delta^{2/3}.
\end{equation}
We consider such values of $\delta$, that   $|f'(y_{z})|>\delta^{2/3}g_{2}g_{1}^{-2/3}/2$. Let us solve the following  equation for $y_{2}$ 
\begin{displaymath}
f_{\delta}(y_{2}) = -\delta^{1/3}
\end{displaymath}
using the Taylor expansion
\begin{displaymath}
f_{\delta}(y)=f_{\delta}(y_{z})+f'_{\delta}(y_{z})(y-y_{z})+\ldots .
\end{displaymath}
The solution reads
\begin{displaymath}
y_{2}= y_{z}+\frac{\delta^{1/3}+\delta^{2/3}g_{1}^{1/3}}{|f_{\delta}'(y_{z})|}.
\end{displaymath}
Trading $f'_{\delta}(y_{z})$ for $f'(y_{z})$ in the above relation changes the result with a term of order $\delta^{1}$, what is negligible. To ensure validity of the solution the reminder of the Taylor expansion $R$ has to be shown irrelevant. It has the form
\begin{displaymath}
R=\frac{f''_{\delta}(s)}{2 {f'_{\delta}}^{2}(y_{z})}(y_{2}-y_{z})^{2},
\end{displaymath}
where $s\in(y_{z},y_{2})$.   The equation (\ref{regul}) does not touch the quantity
\begin{displaymath}
E_{mech}=(n-1)\int_{y_{s}}^{y}\frac{{f'_{\delta}}^{2}(r)}{r} \ dr+ \frac{1}{2} \left({f'_{\delta}}^{2}(y) + f_{\delta}^{2}(y)\right) - \sqrt{f_{\delta}^{2}(y)+\delta^{2}} ,
\end{displaymath}
which is interpreted as the mechanical energy (see section \ref{sect:regul}) at ``time" $y$ plus the energy lost on the ``time" interval $[y_{s},y]$. From this we can  get a bound on $f'_{\delta}$ in terms $f_{\delta}$ for all $y>y_{s}$. Plugging this into the equation (\ref{regul}) a bound for $f_{\delta}''$ is found. Hence, if $\delta$ is small enough, the solution starting with $f_{\delta}(0)>f_{0}$ cannot correspond to a Q-ball profile function as it changes its sign. \\
A crude estimation of $\xi$ allowing the above reasoning gives $\xi \sim \delta^{4/3}$. This is obtained  by finding $y_{z}$ by Taylor expansion of $f$ around $y_{0}$ and by checking the condition $|f'(y_{z})|>\delta^{2/3}g_{2}g_{1}^{-2/3}/2$.\\
The succinct conclusion of this section is
\begin{equation}
\lim_{\delta \rightarrow 0}\hat{f}_{\delta}(0)=f_{0}.
\end{equation}

\subsection{The limit $\hat{f}_{\delta}$ with $\delta\rightarrow 0$}
To investigate the difference between the Q-ball solutions in the
regularized signum-Gordon model and the original one it is
convenient to use the method from the previous section. First, we
denote
\begin{equation}
\hat{\eta}(y)=\hat{f}(y)-\hat{f}_{\delta}(y)
\end{equation}
and for further convenience
\begin{displaymath}
r(\delta)=g_{1}\delta^{2/3}+|f_{0}-\hat{f}_{\delta}(0)|.
\end{displaymath}
For $y<y_{0}$ the equation for $\hat{\eta}$ has the same form as (\ref{etaeq}). As  $\hat{\eta}(0)\neq 0$ the solution of this equation differs slightly  from (\ref{rozw1}), it has the following form
\begin{equation}\label{rozw2}
\hat{\eta}(y)= \int_{0}^{y}G(y,s)s^{n-1}\varphi\left(\hat{f}_{\delta}(s)\right)ds+\frac{f_{0}- \hat{f}_{\delta}(0)}{u_{1}(0)}u_{1}(y).
\end{equation}
Consider  a point $y_{3}$ such that $\hat{f}_{\delta}=\delta^{2/3}\sqrt{2}$. Assume, that
$y_{3}\leq y_{0}$. The term coming from the integration in~(\ref{rozw2}) is not larger
than $g_{1}\delta^{2/3}$, see (\ref{estimd}), (\ref{estim0}) and (\ref{estim}).
As the amplitude of $u_1$ decreases, the second term in the solution is bounded by $|f_{0}-\hat{f}_{\delta}(0)|$. Thus
\begin{displaymath}
\hat{f}(y_{3})\leq \frac{\delta^{2/3}}{\sqrt{2}}+r(\delta).
\end{displaymath}
The two functions are positive, so the difference between them for
any $y$ is  equal to or lesser  than  max\{$\hat{f}(y),
\hat{f}_{\delta}(y)\}$. They are also  decreasing, hence such  bound for
$|\hat{\eta}|$ is valid for all arguments larger  than that one used
in estimation. In that way we obtain the relation
\begin{equation}
|\hat{\eta}(y)|<\frac{\delta^{2/3}}{\sqrt{2}}+r(\delta),
\end{equation}
valid for all $y>0$. If $y_{3}>y_0$, the above estimate remains
valid. To see this, note that  the previous bound for $\hat{\eta}$ holds for all $y<y_{0}$  as $\hat{f}_{\delta}(y_{0})>\delta/\sqrt{2}$. For   $y>y_{0}$ we have the identity $\hat{f}_{\delta}(y)=\hat{\eta}(y)$, hence  $\hat{\eta}$ decreases. 
This completes the proof of one of the main results of this paper: the Q-ball solutions in the regularized signum-Gordon
model approach the solution of the ''sharp" model uniformly. In consequence, the charge and energy computed in regularized model tend to the value known from the original one as $\delta \rightarrow 0$. 

\subsection{The limit $\delta\rightarrow 0$ for energy and charge}\label{DTE}
The numerical results from the section \ref{numerics} point to the agreement between the relation $E(Q)$ in both models of interest. Now we can show, that this is not  an accidental coincidence. 
The fact, that the integration $\int d^{n}x \ f_{\delta}^{2}$ approaches the value known from the signum-Gordon model  follows immediately from the uniform convergence of the functions $\hat{f}_{\delta}$. It is natural to write the  result of the integration in the form $q_{0}+q(\delta)$, where the first term is the limiting value, the second reports on the  $\delta$-dependent corrections. Plugging this into the  original formula for charge (\ref{char}) and trading $\delta$ for the original parameters of the model we get the formula
\begin{equation}
Q=\frac{\pi\lambda^{2}}{\omega^{n+3}}\left(q_{0}+q\left(\frac{2 \epsilon \omega^{2}}{\lambda}\right)\right).
\end{equation}
In the leading order it is the same formula as in the ``sharp" model, the effect of the regularization is negligible both for large charges (small~$\omega$) and tiny regularization parameter $\epsilon$. \\
The same result is true for energy. However, to see this more work is needed. In terms of  $\hat{f}_{\delta}$ the energy functional has the following form
\begin{equation}\label{ener2}
E=\frac{\pi \lambda^{2}}{\omega^{n+2}} \int dr \ r^{n-1} \left[ (\hat{f}_{\delta}{'})^{2} + \hat{f}_{\delta}^{2} + 2\left(\sqrt{\hat{f}_{\delta}^{2}+\delta^{2}}-\delta\right)\right].
\end{equation}
First, the derivative $\hat{f}'_{\delta}$ approaches  $\hat{f}'$; their difference $\hat{\eta}'$ may be bounded  analogously to $\hat{\eta}$. Next, we deal with the potential energy of the field, as the dependence on $\delta$ is explicit there. We separate the $\delta$ independent part in this integral 
\begin{displaymath}
\int dy \ y^{n-1}\left(\sqrt{\hat{f}_{\delta}^{2}+\delta^{2}}-\delta\right)=\int dy \ y^{n-1}\hat{f}_{\delta}-2\delta \int dy \ y^{n-1}\frac{\hat{f}_{\delta}}{\sqrt{\hat{f}_{\delta}^{2}+\delta^{2}}+\delta+\hat{f}_{\delta}}
\end{displaymath}
and show that the  term depending  explicitly on $\delta$ tends to zero. In the above formula the positivity of  $\hat{f}_{\delta}$ is taken into account. Let us split  the area of integration into two parts. First, we treat the integration in a compact  volume
\begin{displaymath}
2 \delta \int_{0}^{y_{0}} dy \ y^{n-1}\frac{\hat{f}_{\delta}}{\sqrt{\hat{f}_{\delta}^{2}+\delta^{2}}+\delta+\hat{f}_{\delta}}<2 \delta \int_{0}^{y_{0}} dy \ y^{n-1}\frac{1}{2}\rightarrow 0.
\end{displaymath}
The integration in the remaining volume is also negligible
\begin{displaymath}
0\leq 2 \delta \int_{y_{0}} dy \ y^{n-1}\frac{\hat{f}_{\delta}}{\sqrt{\hat{f}_{\delta}^{2}+\delta^{2}}+\delta+\hat{f}_{\delta}}<2 \delta \int_{y_{0}} dy \ y^{n-1}\frac{\hat{f}_{\delta}}{2 \delta}\rightarrow \int_{y_{0}} dy \ y^{n-1} \hat{f}=0,
\end{displaymath}
as expected.\\
Denoting the results of integrations in the energy definition  with $\kappa{0}+\kappa(\delta)$ (ana\-lo\-gously to the results of integration in charge definition) we obtain
\begin{equation}
E=\frac{\pi\lambda^{2}}{\omega^{n+2}}\left(\kappa_{0}+\kappa\left(\frac{2 \omega^{2}\epsilon}{\lambda}\right)\right).
\end{equation}
Again, the formula in the leading order is the same as in the model without regularization. This explains the agreement seen in fig. \ref{deqas} - the dependence on regularization parameter $\epsilon$ practically factors out in the relation $E(Q)$.

\section{The absolute stability of Q-balls}
\label{proof}
As  already mentioned, the potential in  (\ref{lag}) does not fall into a class of ``acceptable" ones. For the class Coleman showed in \cite{coleman} that  the Q-ball solutions are absolutely stable, i.e. for a given charge value no configuration can have a  lesser energy. The status of Q-balls  in the regularized signum-Gordon model is at the moment unclear. In this section we are about to adapt  the Coleman's proof  to the theory set by (\ref{lag}). To this end we follow closely his arguments. As originally, our proof is done in three space dimensions.\\
To begin with, we define the Q-ball initial data. A set of initial data is said to be of this type if the spatial distribution of the field is given with a real, positive, spherically symmetric and monotone decreasing to zero function $F$. The condition for time derivative is $\partial_{t}\Phi(t=0,r)=i\omega F(r)$, and $\omega$ is a positive constant. 
The first step in the proof is very general and we just straightforwardly quote it. It states, that for  any set of initial values there exists a set of Q-ball type having the same charge $Q$ and equal or lesser value of energy $E$. As a result we are allowed to constrain the investigation to the energy functional written in the form
\begin{equation}\label{EQstart}
E_{Q}=\int d^{3}x\left[(\nabla F)^{2}+U(F)\right]+\frac{Q^{2}}{I},
\end{equation}
where $F$ is a function giving  the spatial distribution of the initial data of Q-ball type and $I[F]=\int d^{3}x F^{2}$,
\begin{displaymath}
U(F)=\lambda\left(\sqrt{F^{2}+\epsilon^{2}}-\epsilon\right).
\end{displaymath}
 In this form the energy is a function of $F$ and $Q$ is a parameter, $\omega=Q/I$. The aim of the proof is to show, that the minimum of the functional may be reached. 
Before we  proceed, let us discuss the  definition of an ``acceptable" potential. A field potential $U$ is ``acceptable", if
\begin{enumerate}
\item   $U(0)=0$ and $U$ is positive everywhere else. U is twice continuously differentiable, $U'(0)=0$ and $U''(0)=\mu^{2}$.
\item The minimum of $U/F^{2}$ is attained for some $F_{0}\neq 0$.
\item There exist three positive numbers $a$, $b$ and $c>2$, such that
\begin{equation}\label{Colwar}
\frac{1}{2}\mu^{2} F^{2}-U(F)\leq \min({a,b|F|^{c}}).
\end{equation}
\end{enumerate}
The signum-Gordon model spoils all this three conditions. Its regularized version fails to satisfy the second and the third point. As for the second condition, one can say, that the minimum in both models is attained for $\Phi=\infty$.  Instead of this requirement it  suffices, that for some $Q$ there exists a function $F$, for which $\sqrt{2}\mu Q>E$ ( for our convention in Lagrangian  $\sqrt{2}$ appears occasionally). In the regularized model the Q-ball solutions meet this criterion for charges large enough. It follows from the relation $E(Q)$ in the model with ``sharp" potential \cite{ar}
\begin{displaymath}
E\sim Q^{\frac{n+2}{n+3}}.
\end{displaymath}
The meaning of this relation in the regularized potentials is explained in  section~\ref{DTE}.
The third condition for acceptability of the potential is a technical one, useful for some estimates. Happily, we are able to bypass the requirement without any harm to the proof.\\

Let us define:
\begin{displaymath}
K[F]=\int d^{3}x \ \left(\nabla F\right)^{2},
\end{displaymath}
\begin{displaymath}
V[F]=\int d^{3}x \ U(F)=\lambda\int d^{3}x \left(\sqrt{F^{2}+\epsilon^{2}}-\epsilon\right) ,
\end{displaymath}
and 
\begin{displaymath}
W[F]=U[F]-\frac{1}{2}\mu^{2}I[F],
\end{displaymath}
with $\mu^{2}=\lambda/\epsilon$. This quantity satisfies a nice identity
\begin{equation}\label{kwadrat}
W[F]=-\frac{\mu^{2}}{2\lambda^{2}}\int d^{3}x  \ U^{2}(F).
\end{equation}
Hence, $W[F]$ is negative for any $F$. Two decompositions of energy are useful. The first of them is given by (\ref{EQstart}), the second one is
\begin{equation}\label{dEc2}
E_{Q}=K+\frac{\mu^{2}}{2}I+W+\frac{Q^{2}}{I}.
\end{equation}
Now we discuss a meson argument: any spatial distribution of the field vanishing in the infinity may be modified by adding a  function $f$ of compact support
\begin{equation}
F(r)\rightarrow F(r)+L^{-3/2}h\left(\frac{r-d}{L}\right).
\end{equation}
If  $L$ and $d$ are taken large enough the  integrals   $K[F]$ and $W[F]$  stay unaffected and simultaneously $I[F]$  increases by a constant amount. It may be surprising, that this is true for the regularized potential, no matter how small $\delta$ is. This argument however does not work  in case of the ``sharp'' potential. The equation~(\ref{dEc2}) may be written in the form
\begin{equation}\label{EQ}
E_{Q}-K-W=\frac{\mu^{2}}{2}I+\frac{Q^{2}}{I}.
\end{equation}
The r.h.s.  has a minimal value $\sqrt{2}\mu Q$  for $I=\sqrt{2}Q/\mu$. Hence, we can arrange to  add  $\Delta Q$ to the charge value and $\sqrt{2}\mu \Delta Q$ to the energy. 
Consequently if $\inf E_{Q}<\sqrt{2}\mu Q$, there must exist a minimal charge $Q_{min}\geq0$ for which this inequality holds. \\
Consider a sequence of functions $\{F_{i}\}_{i=1}^{\infty}$ for which $lim {E_{Q}[F_{i}]}=inf E$. The existence of such a sequence is guaranteed by the definition of infimum. We can choose $F_{i}$ to  be positive, spherically symmetric and monotone decreasing to zero functions. $K$ is a positive quantity bounded above by the energy. Thus there is a subsequence such that $K$ has a limit. The same reasoning holds for $V$. If $E$, $K$ and $V$ converge, so does also $I$ and  $W$. The limiting values are denoted with tildes, e.g.
\begin{displaymath}
\tilde{K}=\lim_{i\rightarrow \infty}K[F_{i}].
\end{displaymath}
What is more, we can choose such subsequence, that is bounded uniformly in $i$ for all the quantities: $E$, $K$, $V$ and $W$. We assume, this to be done. We will need the inequality
\begin{equation}\label{I}
\tilde{I}>\frac{\sqrt{2}Q}{\mu}.
\end{equation}
Assume, that $\tilde{I}<\sqrt{2}Q/\mu$. Then adding mesons at infinity to the sequence, so that $I=\sqrt{2}Q/\mu$ for almost all $F_{i}$, results in a sequence converging to energy lower than $\tilde{E}=inf{E_{Q}}$, absurdity. If we assume $\tilde{I}=\sqrt{2}Q/\mu$, than $\tilde{W}<-\tilde{K}$ (as $\tilde{E}<\sqrt{2}\mu Q$). The scaling transformation
\begin{displaymath}
F_{i}(y)\rightarrow F_{i}\left(y(1+\alpha)\right),
\end{displaymath}
with small $\alpha$ parameter. Than the energy transforms
\begin{displaymath}
\tilde{E}\rightarrow \tilde{E}-\alpha \tilde{K}-3\alpha\tilde{W}+\ldots,
\end{displaymath}
where the omitted terms are of order $\alpha^{2}$, $\tilde{I}$ is in its stationary point and does not contribute in the first order. Taking $\alpha$ small and negative we could in this way again lower the energy below its infimum.
It is convenient to introduce functions $f_{i}(r)=rF_{i}(r)$, where $r$ is the radial coordinate. With no additional prerequisites we are able to show, that this functions form a uniformly bounded sequence of equicontinious functions. To see this we note, that
\begin{displaymath}
K[F_{i}]=4\pi\int dr \  \left({\frac{d f_{i}}{dr}}\right)^{2}
\end{displaymath}
and
\begin{displaymath}
I[F_{i}]=4\pi\int dr \  f_{i}^{2}.
\end{displaymath}
By Schwarz inequality we obtain 
\begin{equation}\label{kwadrat2}
f^{2}_{i}(r)=-\frac{1}{2}\int_{r}^{\infty} dr \ f_{i} \frac{d f_{i}}{dr}\leq \frac{1}{8 \pi}\sqrt{I[F_{i}]K[I_{i}]}
\end{equation}
and
\begin{equation}
|f_{i}(r_{1})-f_{i}(r_{2})|=\left|\int_{r_{1}}^{r_{2}} dr \frac{d f_{i}}{dr}\right|\leq \sqrt{\frac{[K[F_{i}]|r_{1}-r_{2}|}{8 \pi}}.
\end{equation}
This inequalities legitimate the above statement. Hence, by Ascoli's theorem there exists a subsequence of $\{f_{i}\}$ which is pointwise convergent everywhere and uniformly convergent on any finite interval. This implies the same for $\{F_{i}\}$, except for $r=0$. The limit of the convergent subsequence is denoted with $\tilde{F}$. The task is now to show, that $E_{Q}[\tilde{F}]=\tilde{E}$. \\
$K$ defines a Hilbert - space norm under which the $F$'s are bounded family of vectors. Such a bounded family has always a weakly converging subsequence. The norm of the weak limit is always less than or equal to the limit of norms. Thus,
\begin{equation}\label{esK}
K[\tilde{F}]\leq\tilde{K}.
\end{equation}
Analogously,
\begin{equation}\label{esI}
I[\tilde{F}]\leq\tilde{I}.
\end{equation}
As for $W$ we take two positive numbers $0<r{\_}<r_{+}$ and keeping in mind the relation~(\ref{kwadrat}) and~(\ref{kwadrat2}) we note, that
\begin{displaymath}
\frac{2\pi \mu^{2}}{\lambda^{2}}\int_{0}^{r{\_}} dr \ r^{2} U^{2}(F_{i}) \leq \frac{2\pi \mu^{2}}{\lambda^{2}} \int_{0}^{r{\_}} dr f_{i}^{2}(r)\leq \frac{\mu^{2}}{4 \lambda^{2}}\sqrt{K[F_{i}]I[F_{i}]}r{\_}
\end{displaymath}
and
\begin{displaymath}
\frac{2\pi \mu^{2}}{\lambda^{2}}\int_{r_{+}}^{\infty} dr \ r^{2} U^{2}(F_{i}) \leq \frac{2\pi \mu^{2}}{\lambda^{2}} \int_{r_{+}}^{\infty} dr \ r^{2} F_{i}U(F_{i})\leq \frac{\mu^{2}\sup f_{i}}{2\lambda^{2}r_{+}} V[F_{i}].
\end{displaymath}
Thus,  taking $r{\_}$ and  $r_{+}$ appropriately we can make the above integrals as small as we want. As $F_{i}$ converges uniformly to $\tilde{F}$ in this interval,  we get
\begin{displaymath}
\lim_{i\rightarrow \infty}W[F_{i}]=W[\tilde{F}].
\end{displaymath}
Finally, we show $\tilde{I}=I[\tilde{F}]$. Assume, that $I[\tilde{F}]<\tilde{I}$. Than, by adding meson at infinity we can construct a new function $F'$ such that $W[F']=W[\tilde{F}]$, $K[F']=K[\tilde{F}]$ and $I[F']$ anywhere in between $I[\tilde{F}]$ and $\tilde{I}$. Using (\ref{I}) we can take, that
\begin{displaymath}
\tilde{I}>I[F']>\frac{\sqrt{2}Q}{\mu}.
\end{displaymath}
This implies, that
\begin{displaymath}
\frac{Q^{2}}{I[F']}+\frac{\mu^{2}}{2} I[F']<\frac{Q^{2}}{\tilde{I}}+\frac{\mu^{2}}{2}  \tilde{I}.
\end{displaymath}
Together with equation (\ref{EQ}) it results in a contradiction: $E_{Q}[F']<\tilde{E}$. Thus, $\tilde{I}=I[\tilde{F}]$. By (\ref{dEc2}) and (\ref{esK}) and taking into account the last result we obtain $E_{Q}[\tilde{F}]\leq \tilde{E}$. Since $E_{Q}[\tilde{F}]< \tilde{E}$ is impossible, we conclude, that $E[\tilde{F}]=\tilde{E}$. Having granted the existence of the minimum of the functional, we are legitimate to claim that it corresponds to the solution of the equation
\begin{displaymath}
\frac{\delta E_{Q}[F]}{\delta F}=0,
\end{displaymath}
i.e. the equation (\ref{regul}).\\

The absolute stability  of the Q-balls in the regularized models suggests the  stability of the Q-balls in the signum-Gordon model. It follows from a simple argument. Consider a set of initial data given by a function $F$ to the charge value  $Q$ and the energy $E'$ lower than the energy $E$ of the corresponding Q-ball in the signum-Gordon model. The energy $E$ obeys the  relation (\ref{sigEQ}). We can plug $F$ into the energy functional of an regularized model (\ref{EQstart}) with a parameter~$\epsilon$ and the charge $Q$. As $|F|\geq \sqrt{F^{2}+\epsilon^{2}}-\epsilon$, the energy functional  with any regularization yields then a smaller value than in the case of the ``sharp" potential. For   $\epsilon$ small enough the energy of the corresponding Q-ball may be as close to $E$ as needed, see section~\ref{DTE}.  Hence, for  $\epsilon$ tiny enough the function $F$  results in the energy value smaller than that of the related Q-ball, what has been already proven impossible. 

\section{Conclusions}
We have shown, that the Q-balls are physically relevant solutions of the regularized signum-Gordon model in three spatial dimensions. They are absolutely stable for large values of charge. What is more, we have demonstrated that Q-balls in the regularized  signum-Gordon model approach the solution known from the ``sharp" model. It holds both for profile functions and their global characteristics and is well illustrated by the numerical solutions. For the first time the parabolic approach to the vacuum known in the signum-Gordon model emerged in the limiting procedure. The stability of the solutions in the regularized model guarantees  the stability of the Q-balls in the original model. \\
We have shown, that the regularization does not change some characteristics of the model drastically. We hope that parallel exploration of both  models will shed light on some tough issues, let us mention only the propagation of a perturbation in the model with the ``sharp" potential.\\
Finally, let us point to a very intriguing direction of investigation, i.e. quantization of the models. The question about the role played by the quantum counterparts of Q-balls is both intriguing and hard.

\section{Acknowledgment}
I would like to thank  Henryk Arod\'z for stimulating discussions.

\end{document}